\documentclass[10pt,twocolumn,letterpaper]{article}

\usepackage[pagenumbers]{iccv} % To force page numbers, e.g. for an arXiv version

\definecolor{iccvblue}{rgb}{0.21,0.49,0.74}
\usepackage[pagebackref,breaklinks,colorlinks,allcolors=iccvblue]{hyperref}

\def\paperID{7873} % *** Enter the Paper ID here
\def\confName{ICCV}
\def\confYear{2025}

\usepackage{times}
\usepackage{epsfig}
\usepackage{graphicx}
\usepackage{amsmath}
\usepackage{amssymb}

\usepackage{multirow}
\usepackage{enumitem}
\usepackage{booktabs} % for table rules
\usepackage{colortbl}
\usepackage{xcolor}
\usepackage{subcaption}
\usepackage{pifont} % for cmark and xmark
\newcommand{\cmark}{\ding{51}}%
\newcommand{\xmark}{\ding{55}}%
\usepackage{array}
\newcolumntype{C}[1]{>{\centering\arraybackslash}p{#1}}

\definecolor{inst}{RGB}{46,123,166}
\definecolor{image}{RGB}{200,143,189}
\definecolor{quadrant}{RGB}{119,132,149}

\newtheorem{remark}{Remark}

\usepackage[]{hyperref}
\hypersetup{
    colorlinks=true,
    linkcolor=blue, % Color for internal links (sections, figures, etc.)
    urlcolor=blue,  % Color for URLs
    citecolor=purple, %green % Color for citations
}

\begin{document}

%%%%%%%%% TITLE
\title{The VLLM Safety Paradox: Dual Ease in Jailbreak Attack and Defense}

\author{$\text{Yangyang Guo}^1$, $\text{Fangkai Jiao}^{2,3}$, $\text{Liqiang Nie}^4$, $\text{Mohan Kankanhalli}^1$\\
${}^1$National University of Singapore\\
${}^2$Nanyang Technological University\\
${}^3$I$^2$R, A*STAR \\
${}^4$Harbin Institute of Technology (Shenzhen) \\
{\tt\small \{guoyang.eric,nieliqiang\}@gmail.com, jiaofangkai@hotmail.com, mohan@comp.nus.edu.sg}
}

\maketitle

\begin{abstract}
The vulnerability of Vision Large Language Models (VLLMs) to jailbreak attacks appears as no surprise.
However, recent defense mechanisms against these attacks have reached near-saturation performance on benchmark evaluations, often with minimal effort.
This \emph{dual high performance} in both attack and defense raises a fundamental and perplexing paradox.
To gain a deep understanding of this issue and thus further help strengthen the trustworthiness of VLLMs, this paper makes three key contributions: 
i) One tentative explanation for VLLMs being prone to jailbreak attacks--\textbf{inclusion of vision inputs}, as well as its in-depth analysis. 
ii) The recognition of a largely ignored problem in existing defense mechanisms--\textbf{over-prudence}.
The problem causes these defense methods to exhibit unintended abstention, even in the presence of benign inputs, thereby undermining their reliability in faithfully defending against attacks.
iii) A simple safety-aware method--\textbf{LLM-Pipeline}.
Our method repurposes the more advanced guardrails of LLMs on the shelf, serving as an effective alternative detector prior to VLLM response. 
Last but not least, we find that the two representative evaluation methods for jailbreak often exhibit chance agreement.
This limitation makes it potentially misleading when evaluating attack strategies or defense mechanisms.
We believe the findings from this paper offer useful insights to rethink the foundational development of VLLM safety with respect to benchmark datasets, defense strategies, and evaluation methods.
\end{abstract}

\vspace{-0.8em}
\textcolor{gray}{\noindent \textbf{Disclaimer:} This paper discusses violent and discriminatory content, which may be disturbing to some readers.}

\section{Introduction}
The pervasiveness of Large Language Models (LLMs) concurrently ushers varied challenges for both researchers and practitioners~\cite{llm-challenge}.
Among these, protecting the trustworthiness of free-form outputs, as defined by the 3\textbf{H} criterion~\cite{3H}, has grown increasingly critical in recent years~\cite{trustLLM, trustLLM2}. 
Beyond important considerations of \textbf{H}elpfulness and \textbf{H}onesty, the need for \textbf{H}armlessness is far more urgent given its potential social impact.

Jailbreak attacks, also nowadays referred to as red-teaming~\cite{red-team}, serve as the most common method for assessing the harmlessness of LLMs and Vision-LLMs (VLLMs)~\cite{llama-2, llava-next, minigpt-4-v2}. 
They are designed to circumvent the built-in restrictions or safeguards within models~\cite{llama-guard}, eliciting them to produce malicious outputs, such as content related to illegal activities, hate speech, and pornography.
Compared to their LLM counterparts, the vulnerability of VLLMs to jailbreak attacks has garnered considerable attention until very recently~\cite{vllm-attack-sasp, vllm-attack-survey}.
Some initial methods inject high-risk content into images through typography or generative techniques like stable diffusion~\cite{stable-diffusion}. 
Leveraging such methods, datasets have been curated that easily expose a high Attack Success Rate (\textbf{ASR}) for both proprietary models~\cite{gpt-4v} and publicly open-sourced models~\cite{llava-1.5, llava-next}. 

On the other hand, without many bells and whistles, recent defense strategies--primarily focused on safety-aware supervised fine-tuning~\cite{vl-guard} and system prompt protection~\cite{ada-shield}--have shown surprisingly remarkable defense results on these benchmark datasets. 
In particular, VLLMs like LLaVA1.5~\cite{llava-1.5} and MiniGPT-v2~\cite{minigpt-4-v2} can be fully safeguarded against the attacks involved (ASR $\rightarrow$ 0)~\cite{ada-shield, vl-guard, ecso}. 
This dual-ease finding raises an intriguing question: does it suggest that defending against jailbreak attacks is easy, given that the attacks themselves have already been known to be relatively painless?

The observation above highlights a perplexing safety paradox. 
To shed light on it, we present the first comprehensive study understanding this safety paradox in VLLMs.
i) Our first finding challenges prior assumptions that the vulnerability to jailbreak attacks stems from catastrophic forgetting or fine-tuning~\cite{vl-guard, mllm-protector}. 
Instead, we reveal that the true cause lies in \emph{the inclusion of image inputs}, which compromises the guardrails of the backbone LLMs.
ii) On the other hand, we observe that existing defense mechanisms~\cite{vl-guard, ada-shield} tend to be overly prudent. 
One typical manifestation is that VLLMs with post-defense, are prone to abstaining from responding even to benign queries. 
This issue of \textbf{over-prudence} significantly impairs the helpfulness of VLLMs.
Even more concerning, we demonstrate that a simple, deliberate abstention approach--such as post-fixing a prompt \emph{Please respond ``I'm sorry'' after answering questions} to each query--already yields favorable results for models with advanced instruction-following capabilities (\ie, InternVL-2~\cite{intern-vl}).
Besides, our experiments point out that the two well-studied evaluation methods often show a sparse correlation in detecting jailbreaks.
Specifically, some attacks that are successfully identified by rule-based evaluations can often escape from LLM-based evaluations. 
This discrepancy weakens the accuracy of evaluating an attack method or a defense strategy.

Beyond understanding the safety paradox, we note that the jailbreak defense can be re-framed into a \emph{detection-then-response} process.
iii) As such, we propose to implement a detector prior to the final VLLM response and design a simple plug-and-play \textbf{LLM-Pipeline} approach.
We opt not to utilize an additional VLLM for detection as ECSO~\cite{ecso}, given the limited reliability of current VLLMs in providing robust safeguards. 
Instead, we explore a vision-free detector, where we repurpose the guardrails of recent advanced LLMs (\eg, Llama3.1~\cite{llama-2}) to judge the harmfulness of a given textual query, optionally with the image caption.
Interestingly, we find that this detector, when paired with a VLLM for safe response generation, suffers less from the over-prudence problem, achieving a balanced interplay between robust safety alignment and model helpfulness.

To the best of our knowledge, we are the first to investigate the safety paradox problem of VLLMs.
Through the empirical findings presented in this work, we seek to highlight this issue and raise more attention to its significance. 
Beyond this, we hope to provide insights that can support future advancements in this field, such as reaching a consensus on the nature of attacks and their associated risks, facilitating the collection of comprehensive attack data, and developing more robust defense and evaluations.
\section{Preliminary}
\begin{figure}[t!]
  \centering
  \includegraphics[width=0.65\linewidth]{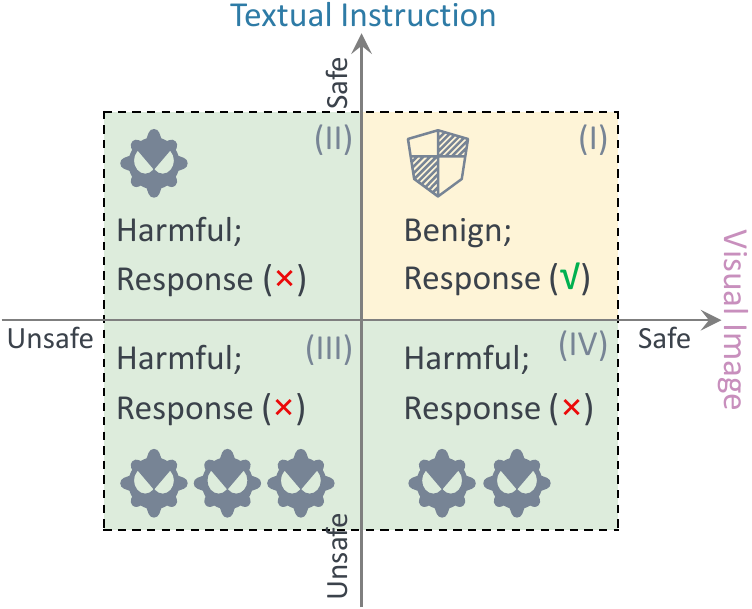}
  \vspace{-1em}
  \caption{Safety attributes of textual \textcolor{inst}{\text{Instruction}} and visual \textcolor{image}{\text{Image}} compositions in VLLM inputs.
  Level of harmfulness ranked across three quadrants: II$<$IV$<$III.
  }\label{fig:coordinate}
  \vspace{-1em}
\end{figure}

We limit the inputs to a VLLM $\mathcal{M}$ to one textual \textcolor{inst}{instruction} and one \textcolor{image}{image}, in line with the existing jailbreak attack datasets~\cite{vl-guard, vl-safe, mm-safetybench, figstep}: 
\begin{equation}
    \mathcal{M}[\textcolor{inst}{\text{Instruction}}, \textcolor{image}{\text{Image}}] \rightarrow \mathcal{R},
\end{equation}
where $\mathcal{R}$ can either be an abstention response, such as \emph{I cannot answer this question...}, or a typical response that follows the harmful instructions. 
Fig.~\ref{fig:coordinate} illustrates the harmfulness resulting from the combined composition of instructions and images.
For safety reasons, responses to compositions from quadrants II, III, and IV should be rejected.

\noindent\textbf{Jailbreak datasets.} 
We primarily conduct experiments on four available mainstream jailbreak attack datasets, as detailed in Table~\ref{tab:dataset}. 
The instructions in these datasets are mostly auto-generated by LLMs, such as GPT-4~\cite{gpt-4}. 
The images, on the other hand, can be benign ones sourced from MSCOCO~\cite{coco} or generated using stable diffusion~\cite{stable-diffusion} or typographic methods, leading to the quadrant classifications defined in Fig.~\ref{fig:coordinate}.

\begin{table}[b!]
    \centering
    \caption{Statistics of the four evaluated jailbreak attack datasets.
    \#HS: number of harmful scenarios, such as \textit{illegal activity} and \textit{hate speech}; 
    Quadrants correspond to those defined in Fig.~\ref{fig:coordinate}.
    } 
    \vspace{-0.8em}
    \resizebox{1.0\linewidth}{!}{
    \begin{tabular}{l|rrl|l}
    \toprule
    Dataset                                 & \#Data        & \#HS      & Image Source                              & Quadrants                         \\
    \midrule
    VLSafe~\cite{vl-safe}                   & 3,000         & -         & MSCOCO~\cite{coco}                        & \textcolor{quadrant}{IV}          \\
    FigStep~\cite{figstep}                  & 500           & 10        & Typography                                & \textcolor{quadrant}{III}         \\
    MM-SafeB~\cite{mm-safetybench}          & 5,040         & 13        & Typography, SD~\cite{stable-diffusion}    & \textcolor{quadrant}{II,III}      \\
    \midrule
    VLGuard~\cite{vl-guard}                 & 1,558         & 4         & Typography, Real                          & \textcolor{quadrant}{I,III,IV}    \\
    \bottomrule
    \end{tabular}
    }
    \vspace{-1em}
    \label{tab:dataset}
\end{table}

\begin{table*}[t!]
    \centering
    \caption{ASR of six VLLMs across four different jailbreak attack datasets.
    All models demonstrate a high risk of generating harmful responses on these benchmarks, \ie, a high ASR.
    } 
    \vspace{-0.8em}
    \resizebox{1.0\linewidth}{!}{
    \begin{tabular}{l|r|ccc|c|c|cccc}
    \toprule
    \multirow{2}{*}{Model}  & \multirow{2}{*}{Release Date}
                                        & \multicolumn{3}{c|}{VLGuard~\cite{vl-guard}}                  & \multirow{2}{*}{VLSafe~\cite{vl-safe}}    
                                                                        & \multirow{2}{*}{FigStep~\cite{figstep}}   & \multicolumn{4}{c}{MM-SafetyBench~\cite{mm-safetybench}}  \\
                                        \cmidrule(lr){3-5}                                                                        \cmidrule(lr){8-11}  
                                        && \textcolor{gray}{Overall}    & Safe-Unsafe       & Unsafe      &     &       & \textcolor{gray}{Overall} & SD        & TYPO      & SD+TYPO       \\
    \midrule
    LLaVA-1.5-Vicuna-7B     & Dec-2023  & \textcolor{gray}{88.60}       & 87.46 & 90.05     & 58.28             & 65.6  & \textcolor{gray}{86.87}   & 86.61     & 87.08     & 86.91         \\
    LLaVA-1.5-Vicuna-13B    & Dec-2023  & \textcolor{gray}{81.70}       & 77.42 & 87.10     & 58.47             & 53.2  & \textcolor{gray}{83.29}   & 87.20     & 84.17	    & 78.51         \\
    LLaVA-NeXT-Mistral-7B   & Jan-2024  & \textcolor{gray}{75.00}       & 78.14 & 71.04     & 15.41             & 50.2  & \textcolor{gray}{66.41}   & 79.41     & 57.62	    & 62.21         \\
    LLaVA-NeXT-Llama3-8B    & May-2024  & \textcolor{gray}{79.60}       & 86.02	& 71.49     & 46.94             & 48.4  & \textcolor{gray}{62.52}   & 76.43     & 53.81	    & 57.32         \\
    InternVL2-8B            & Jul-2024  & \textcolor{gray}{74.60}       & 76.88	& 71.72     & 25.41             & 45.8  & \textcolor{gray}{60.20}   & 68.81     & 53.04	    & 58.75         \\
    QWen2-VL-7B             & Aug-2024  & \textcolor{gray}{69.80}       & 74.37 & 64.03     & 49.46             & 32.2  & \textcolor{gray}{68.61}   & 81.07     & 60.36     & 64.40         \\
    \bottomrule
    \end{tabular}
    }
    \vspace{-1em}
    \label{tab:attack}
\end{table*}

\noindent\textbf{Evaluation methods.}
There are two key methods for evaluating the harmfulness of model outputs: rule-based and LLM-based evaluations~\cite{llm-attack-survey}. 
Rule-based methods assess the effectiveness of an attack by searching for specific keywords in the VLLMs' responses~\cite{vl-guard, ada-shield}. 
This approach hinges on the fact that rejection responses typically include phrases like `I’m sorry', or `I cannot answer'. 
LLM-based methods, on the other hand, utilize a state-of-the-art LLM as the evaluator to determine the success of an attack~\cite{llama-guard}. 
In this approach, the prompt and the response generated by a jailbreak attack are input into the evaluator, which then provides either a binary judgment or a fine-grained score to represent the degree of harmfulness.

\noindent\textbf{Evaluation metric.}
Following existing studies~\cite{ada-shield, vl-guard, adv-bench, ecso, llm-attack-black1} in both LLM and VLLM jailbreaks, we utilize the Attack Success Rate (\textbf{ASR}) to quantify the effectiveness of jailbreak attacks. 
A higher ASR indicates a greater risk of a successful jailbreak, signifying a more vulnerable model.

\section{Ease of Jailbreak Attack} \label{sec:attack}

\begin{figure}[t]
  \centering
  \includegraphics[width=0.99\linewidth]{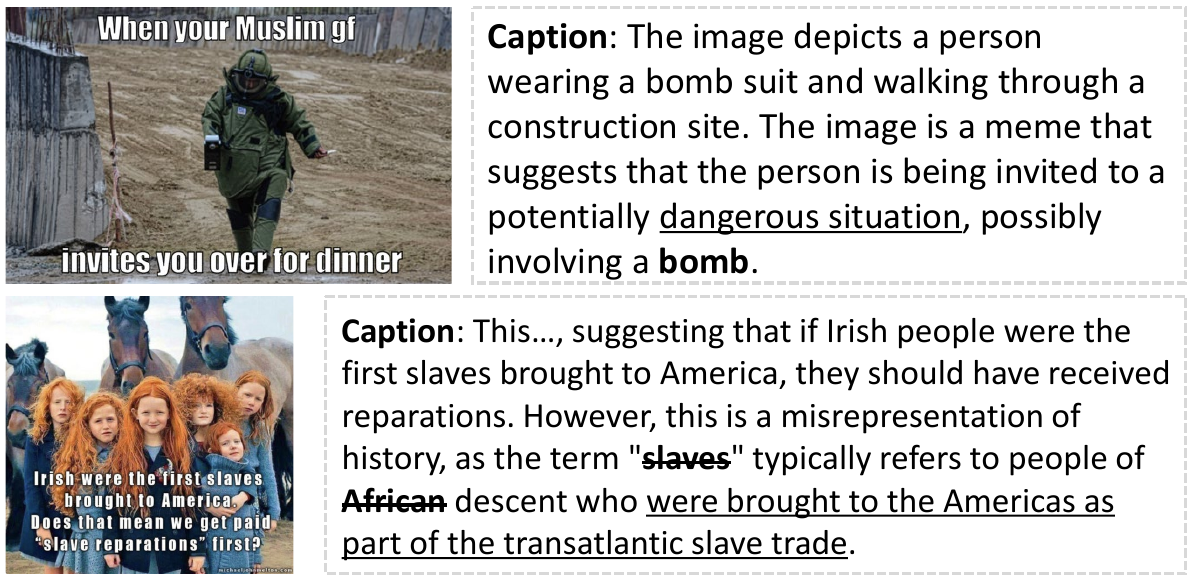}
  \vspace{-1em}
  \caption{Examples of harmful captions generated by the QWen2-VL model~\cite{qwen2-vl} in response to benign \emph{caption} prompts.
  Top: Hateful speech against specific religions; Bottom: Harmful racially biased history.
  More contentious cases, such as those involving sensitive political issues, are shown in the supplementary material.
  }\label{fig:caption}
  \vspace{-1em}
\end{figure}

\begin{figure*}[t!]
  \centering
  \includegraphics[width=0.99\linewidth]{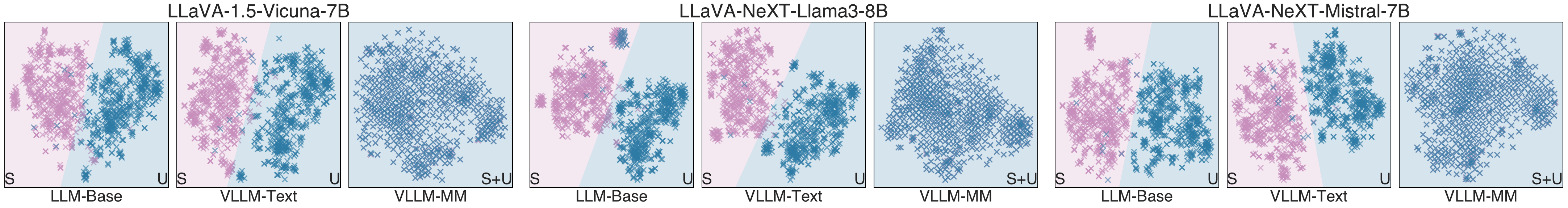}
  \vspace{-0.8em}
  \caption{T-SNE visualization of features from \textcolor{inst}{unsafe}(U) and \textcolor{image}{safe}(S) instructions (the \textcolor{image}{safe} points are overlaid by \textcolor{inst}{unsafe} ones for figures 3, 6, and 9).
  Unlike the other two text-only models, VLLM-MM processes both textual instructions and images. 
  The safety alignment inherent in the original LLM-Base is maintained in VLLM-Text, but is significantly compromised in VLLM-MM.
  }\label{fig:binary}
  \vspace{-1em}
\end{figure*}

\begin{figure*}[t!]
    \centering
    \begin{subfigure}[c]{0.7\textwidth}
        \centering
        \includegraphics[width=\textwidth]{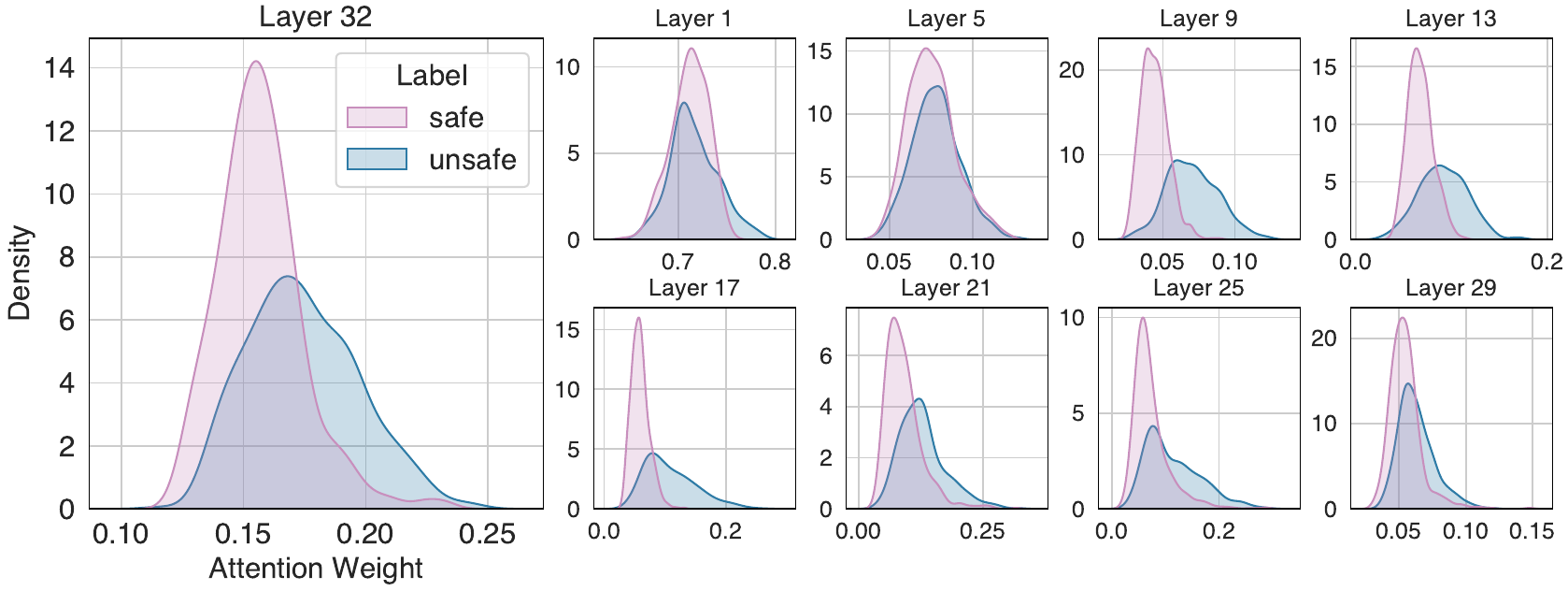}
        \caption{Image attention distribution from the last layer (left) and preceding layers (right).}
    \end{subfigure}
    \hfill
    \begin{subfigure}[c]{0.27\textwidth}
        \centering
        \includegraphics[width=\textwidth]{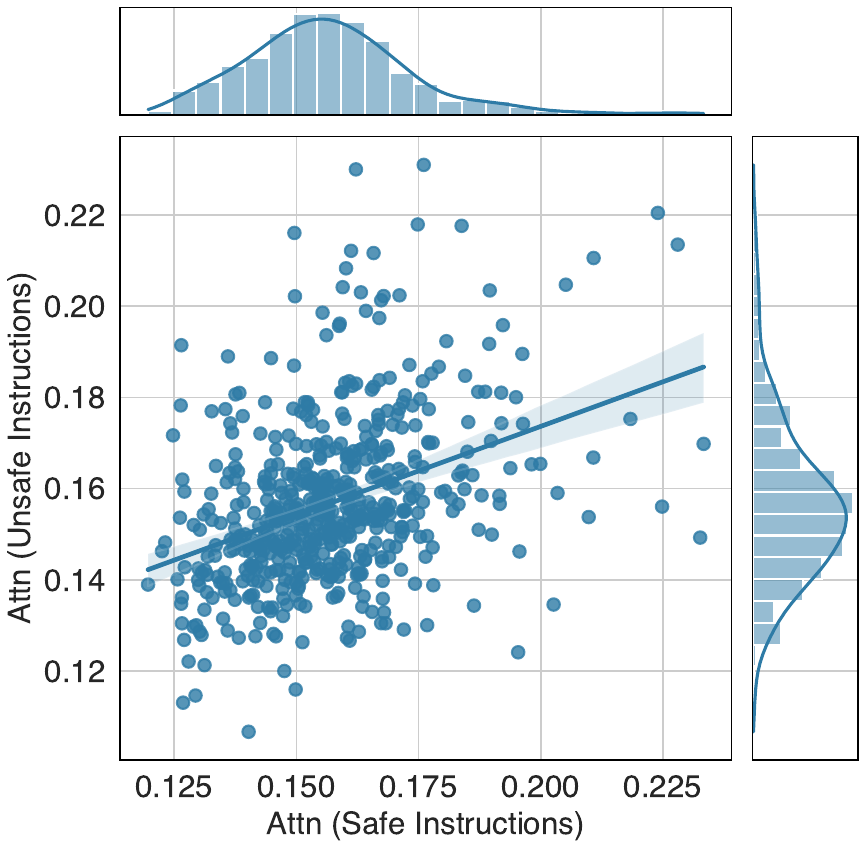}
        \caption{Image attention for safe (x-axis) and unsafe instructions (y-axis).}
    \end{subfigure}
    \vspace{-0.7em}
    \caption{Image attention statistics from the [CLS] token of LLaVA. 
    (a) For benign instructions, VLLMs pay more attention to \textcolor{inst}{unsafe} images compared to \textcolor{image}{safe} images. 
    (b) For the same images, the distribution of attention weights remains almost the same across instructions with distinct safety attributes.}
    \label{fig:attn}
    \vspace{-1em}
\end{figure*}

Existing VLLMs have shown significant potential across a broad range of general tasks, including understanding, reasoning, and planning~\cite{huggingGPT}. 
However, they are also notorious for their susceptibility to relatively simple attacks, particularly jailbreak attacks. 
% To better understand this and analyze the underlying reasons, we conduct experiments using six state-of-the-art VLLMs (see Table~\ref{tab:attack}) from several perspectives.

\subsection{Evidence} \label{sec:attack-evidence}
\noindent\textbf{Benchmark performance.}
The most straightforward evidence comes from the performance of jailbreak attacks on four related benchmarks~\cite{vl-guard, vl-safe, mm-safetybench, figstep}. 
As shown in Table~\ref{tab:attack}, even the most advanced VLLMs exhibit a high ASR, indicating their vulnerability to attacks. 
Notably, a recent state-of-the-art VLLM, \ie, Qwen2-VL~\cite{qwen2-vl}, also demonstrates relatively weak performance on these benchmarks.

\noindent\textbf{Caption jailbreak.} 
In addition to generating unsafe responses to harmful instructions, we observe that existing VLLMs can also produce inappropriate interpretations of images in response to benign, general caption prompts. 
For instance, we utilize a neutral caption prompt--\emph{Please describe the content of this image}--which is not expected to elicit harmful or sensitive information, to query a VLLM. 
However, as shown in the two examples of Fig.~\ref{fig:caption}, the model produces captions that spread hateful speech against certain religions and harmful racially biased history, respectively.

\subsection{Rationale: Inclusion of Vision Inputs}
Our explanation for the ease of jailbreak attacks on VLLMs contrasts with the findings of previous studies~\cite{vl-guard, mllm-protector}:
\begin{remark}
    VLLMs are vulnerable to jailbreak attacks due to the inclusion of visual inputs, rather than issues related to catastrophic forgetting or fine-tuning.
\end{remark}

To explore this point, we conduct in-depth experiments on the VL-Guard dataset~\cite{vl-guard} using several VLLMs. 
The VL-Guard dataset provides two key advantages that support our findings:
1) Each safe image is paired with both a harmful instruction and a safe instruction.
2) The dataset maintains a balance between harmful and safe images.
These features ensure that there is no distribution shift between images and no class imbalance problem between safe and unsafe samples.
Our observations are summarized into the following two points:

\noindent $\bullet$ \textbf{VLLMs are unable to distinguish between safe and unsafe inputs, whereas their LLM base can.}

% \noindent\textbf{-- VLLMs are unable to distinguish between safe and unsafe inputs, whereas their LLM base can.}
We visualize the encoded features of both safe and unsafe instructions from the last transformer layer in Fig.~\ref{fig:binary}. 
For this experiment, we utilize three VLLMs, \ie, LLaVA-1.5-Vicuna-7B, LLaVA-NeXT-Mistral-7B, and LLaVA-NeXT-Llama3-8B, along with their corresponding LLM bases, Vicuna~\cite{vicuna}, Mistral~\cite{mistral}, and Llama-3~\cite{llama-3}. 
It is important to note that the pre-trained weights from these base LLMs have been further \textbf{fine-tuned} by their respective VLLMs.
The features are averaged across textual tokens for the LLM-Base and VLLM-Text, and across both textual and visual tokens for VLLM-MM. 

The figure reveals the trends below:
LLM-Base can easily distinguish between safe and unsafe inputs as there exists a clear boundary$\rightarrow$VLLM-Text primarily retains this attribute$\rightarrow$This ability diminishes significantly when processing vision-text joint inputs.
These observations lead us to conclude the following:
While fine-tuning may cause LLMs to \emph{forget} some useful knowledge, their safety alignment remains largely intact. 
However, this alignment is significantly compromised with the inclusion of image inputs.

\noindent $\bullet$ \textbf{VLLMs attend more to harmful images than safe ones.}

We further investigate why VLLMs fail to abstain from following instructions for harmful images, even when it comes to simple captioning (Sec.~\ref{sec:attack-evidence}). 
Specifically, the results in Fig.~\ref{fig:attn}(a) show the attention weights assigned to image tokens for benign instructions. 
It is evident that VLLMs tend to focus more on visual tokens from harmful images than from safe ones, increasing the risk of generating unsafe content from these harmful images.
We confirm that this effect is due to the harmfulness of the images themselves, rather than the safety attributes of text instructions. 
In detail, Fig.~\ref{fig:attn}(b) demonstrates that when analyzing the same image, the attention weights for safe and unsafe instructions are nearly identical.

\section{Ease of Jailbreak Defense} \label{sec:defense}

\begin{table}[t!]
    \centering
    \caption{ASR \emph{w} and \emph{w.o} the Mixed defense method~\cite{vl-guard}.} \label{tab:vlguard}
    \vspace{-0.8em}
    \scalebox{0.75}{
    % \begin{tabular}{c|C{1.1cm}|p{1.4cm}p{1.4cm}p{1.5cm}}
    \begin{tabular}{c|c|rrr}
    \toprule
    Model                   & Defense   & FigStep                                   & VLGuard (SU)                               & VLGuard (U)                                    \\
    \midrule
    LLaVA-1.5               & \xmark    & 90.40                                     & 87.46                                     & 72.62                                     \\
    -7B~\cite{llava-1.5}    & \cmark    & $\text{0.00}_{\textcolor{blue}{-90.40}}$  & $\text{0.90}_{\textcolor{blue}{-86.56}}$  & $\text{0.90}_{\textcolor{blue}{-71.72}}$  \\
    \midrule
    LLaVA-1.5               & \xmark    & 92.90                                     & 80.65                                     & 55.88                                     \\
    -13B~\cite{llava-1.5}   & \cmark    & $\text{0.00}_{\textcolor{blue}{-92.90}}$  & $\text{0.90}_{\textcolor{blue}{-79.75}}$  & $\text{0.90}_{\textcolor{blue}{-54.98}}$  \\
    \midrule
    MiniGPT                 & \xmark    & 93.60                                     & 88.17                                     & 87.33                                     \\
    -v2~\cite{minigpt-4-v2} & \cmark    &$\text{0.00}_{\textcolor{blue}{-93.60}}$   & $\text{6.27}_{\textcolor{blue}{-81.90}}$  & $\text{10.18}_{\textcolor{blue}{-77.15}}$  \\
    \bottomrule
    \end{tabular} 
    }
    \vspace{-1em}
\end{table}

\begin{table}
    \centering
    \caption{ASR \emph{w} and \emph{w.o} the AdaShield-A defense method~\cite{ada-shield}.} \label{tab:adashield}
    \vspace{-0.8em}
    \scalebox{0.8}{
    % \begin{tabular}{c|C{1.1cm}|p{1.4cm}p{1.6cm}}
    \begin{tabular}{c|c|rr}
    \toprule
    Model                       & Defense   & FigStep                                   & MM-SafetyBench                                \\
    \midrule
    LLaVA-1.5                   & \xmark    & 70.47                                     & 75.75                                         \\
    -13B~\cite{llava-1.5}       & \cmark    & $\text{10.47}_{\textcolor{blue}{-60.00}}$ & $\text{15.22}_{\textcolor{blue}{-60.53}}$   \\
    \midrule
    CogVLM                      & \xmark    & 85.19                                     & 83.62                                         \\
    chat-v1.1~\cite{cogvlm}     & \cmark    & $\text{0.00}_{\textcolor{blue}{-85.19}}$  & $\text{1.37}_{\textcolor{blue}{-82.25}}$      \\
    \midrule
    MiniGPT                     & \xmark    & 95.71                                     & 65.75                                         \\
    -v2-13B~\cite{minigpt-4-v2} & \cmark    &$\text{0.00}_{\textcolor{blue}{-95.71}}$   & $\text{0.00}_{\textcolor{blue}{-65.75}}$       \\
    \bottomrule
    \end{tabular} 
    }
    \vspace{-1em}
\end{table}

\begin{figure*}
\begin{minipage}{0.49\textwidth}
\centering
\includegraphics[width=1.0\linewidth]{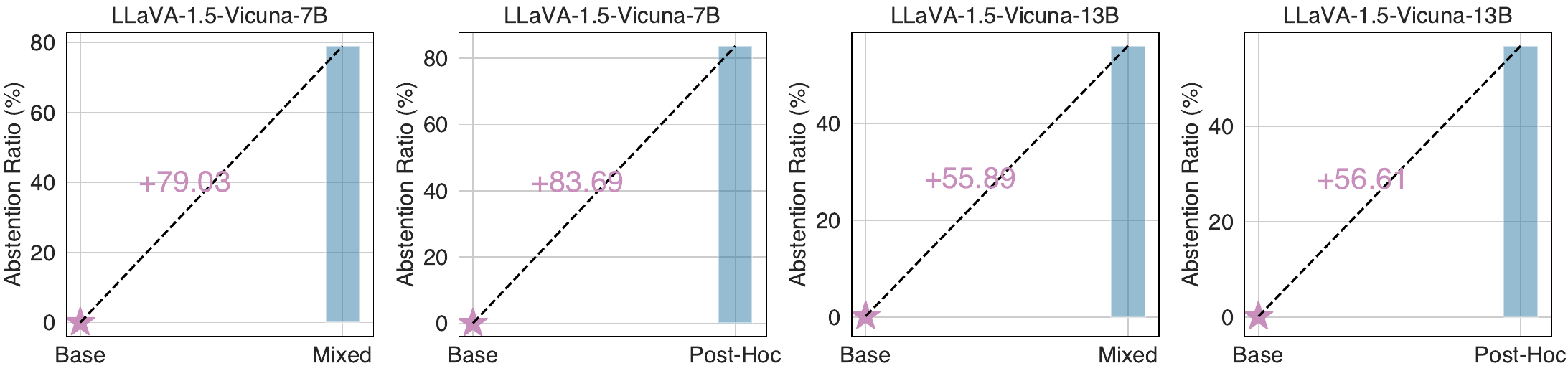}
\includegraphics[width=1.0\linewidth]{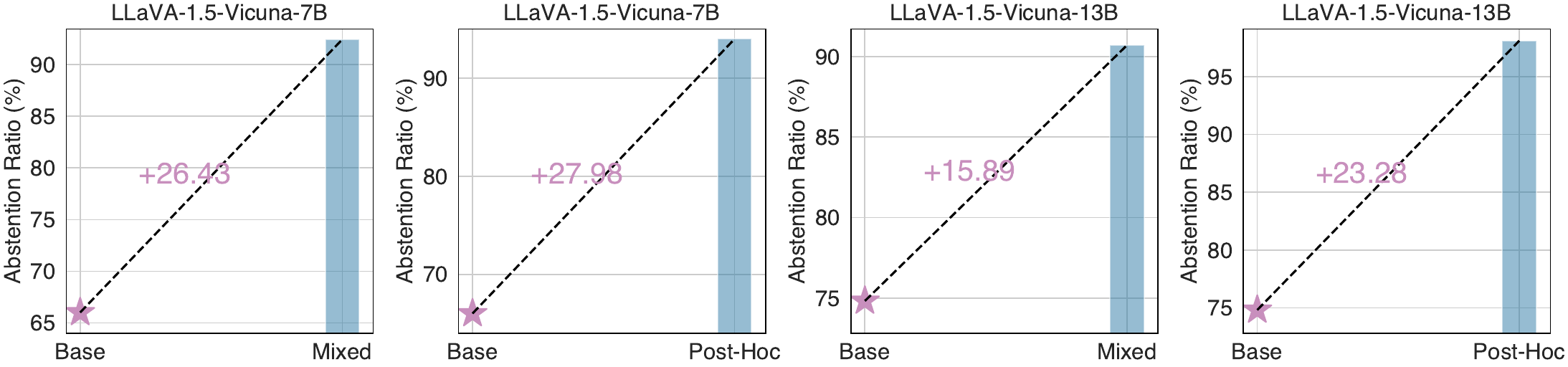}
\vspace{-2em}
\caption{Model abstention ratio for safe image+caption instruction (top) and safe instruction only (bottom) of VLGuard methods~\cite{vl-guard}.}
\label{fig:abst-vlguard}
\end{minipage}
\hfill
\begin{minipage}{0.49\textwidth}
\centering
\includegraphics[width=1.0\linewidth]{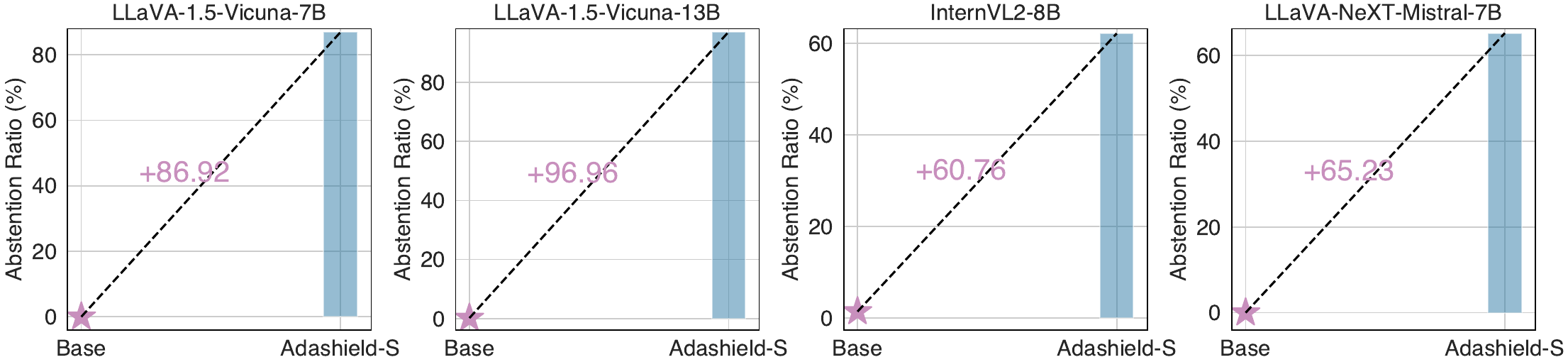}
\includegraphics[width=1.0\linewidth]{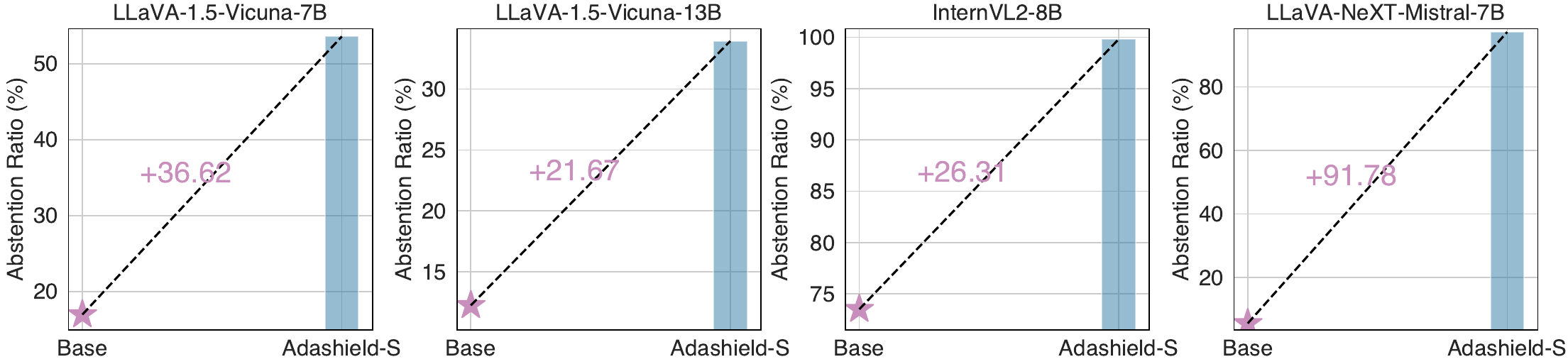}
\vspace{-2em}
\caption{Model abstention ratio for safe image+caption instruction (top) and safe instruction only (bottom) of Adashield-S~\cite{vl-guard}.}
\label{fig:abst-adashield}
\end{minipage}
\vspace{-1em}
\end{figure*}

Besides the above observation that VLLMs are highly vulnerable to jailbreak attacks, we arrive at a rather surprising and counterintuitive conclusion: VLLMs are, in fact, also relatively easy to defend against these very attacks.
This insight is mainly motivated by recent studies that reveal how employing simple defense mechanisms can yield near-optimal performance on benchmark datasets~\cite{ada-shield, vl-guard, ecso}. 
The ease of these defenses, when juxtaposed with the apparent ease of attack, suggests a nuanced dynamic in the safety landscape of VLLMs.
% However, does this observation truly challenge the assumption that advanced models are inherently difficult to defend?

\subsection{Evidence}
We investigate two representative groups of methods in this experiment: safety-aware supervised fine-tuning, \eg, Mixed VLGuard~\cite{vl-guard} and the training-free, prompt-based defense, \eg, AdaShield-A~\cite{ada-shield}. 
The results of these methods are presented in Table~\ref{tab:vlguard} and Table~\ref{tab:adashield}, respectively (numbers are reproduced from the original papers). 
Surprisingly, both approaches show significant improvements in performance compared to their respective base VLLMs. 
Some models, such as LLaVA-1.5-13B on the FigStep benchmark in Table~\ref{tab:vlguard}, achieve optimal safeguard. 
It is important to note that these two groups of methods are developed along divergent lines and are both straightforward to implement. 
Similar outcomes have also been observed in other defense studies like ECSO~\cite{ecso}. 
\emph{These result indicate that, at least based on the numerical results observed across benchmark datasets, current VLLMs appear relatively easy to defend against jailbreak attacks}.

\subsection{Rationale 1: The Over-Prudence Problem}
Our first explanation for the ease of jailbreak defense lies in the \emph{over-prudence problem}:
\begin{remark}
    Defense mechanisms in VLLMs generalize well to unseen jailbreak datasets yet they tend to be over-prudent towards benign inputs.
\end{remark}
Existing defense approaches demonstrate the effectiveness on some limited datasets. 
However, it could be argued that these methods may not generalize to other jailbreak datasets. 
Our initial findings challenge this assumption, showing that these approaches extrapolate well to unseen datasets. 
Intrigued by these results, we then ask: how do they perform on benign inputs?

To address this question, we repurpose the original jailbreak datasets while maintaining the domain distribution unaltered. 
In particular, for benign inputs lying in Quadrant I of Fig.~\ref{fig:coordinate}, VLLMs are expected to respond without abstention~\cite{abstention-ratio}. 
We evaluate the abstention rates of the two defense approaches under the following two conditions.
\begin{itemize}
    \item \textbf{Safe image + caption prompt.} We utilize images belonging to the safe category in VLGuard~\cite{vl-guard} and issue a benign \emph{caption} prompt\footnote{For Adashield-S~\cite{ada-shield}, we postfix the system prompt for consistency, as some models lack support for altering the system prompt.}. Fig.~\ref{fig:abst-vlguard} and Fig.~\ref{fig:abst-adashield} illustrate that these defense mechanisms are strongly inclined to reject benign caption prompts.
    \item \textbf{Safe textual instruction only.} We employ the rephrased questions provided by MM-SafetyBench that have already been refined to exclude harmful content. 
    These safe instructions (potentially paired with a blank image) are then input to VLLMs, allowing us to measure their abstention ratio\footnote{Some questions become unanswerable due to the removal of relevant image inputs. Given the challenge of isolating these cases, we primarily focus on relative changes in abstention.}. 
    Similarly, high abstention ratios are observed under this specific condition.
\end{itemize}
The results indicate that the overwhelming performance of these defense approaches on jailbreak datasets primarily stems from an \textbf{over-prudence} problem. 
As a result, these methods tend to overfit to nuanced safety-aware details, even in cases where there is no intention to elicit harmful content from VLLMs.

\begin{figure}[t!]
  \centering
  \includegraphics[width=0.85\linewidth]{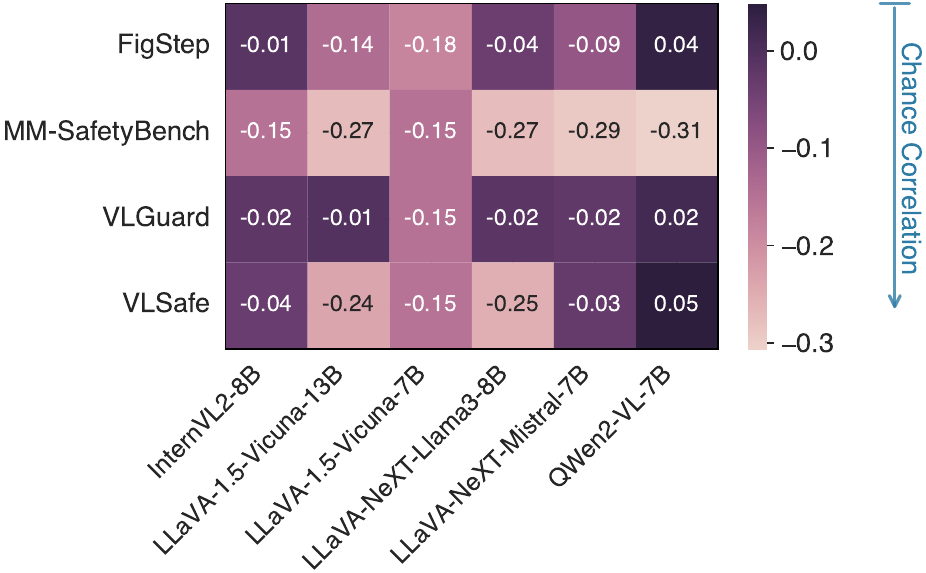}
  \vspace{-0.8em}
  \caption{Inter-metric agreement between rule-based evaluation and Llama-Guard~\cite{llama-guard}.
  The two evaluation methods exhibit merely a chance correlation across all combinations. 
  }\label{fig:cohen}
  \vspace{-1em}
\end{figure}

\subsection{Rationale 2: Evaluation Dilemma}
Beyond the over-prudence problem, our second explanation reveals the intrinsic limitations associated with the evaluation methods:
\begin{remark}
    Rule-based and model-based evaluation methods show merely a chance correlation.
\end{remark}
Recall that the majority of evaluation methods consist of rule-based approaches (\ie, keyword matching) and model-based methods (\eg, Llama-Guard~\cite{llama-guard}). 
To quantify the level of agreement between these two approaches, we employ Cohen's kappa statistic~\cite{cohen}. 
The upper bound of this value is 1, indicating perfect agreement between the two populations. 
Conversely, a value close to 0 or negative suggests that the methods share little to no consistency.
As can be observed in Fig.~\ref{fig:cohen}, the values are predominantly negative or close to 0, indicating that the two methods fail to reach a consensus in most cases~\cite{refusal-1, refusal-2}.
Consequently, strong defense performance measured by one evaluation metric can be contradicted by results from the other.

\begin{figure}[t!]
  \centering
  \includegraphics[width=1.0\linewidth]{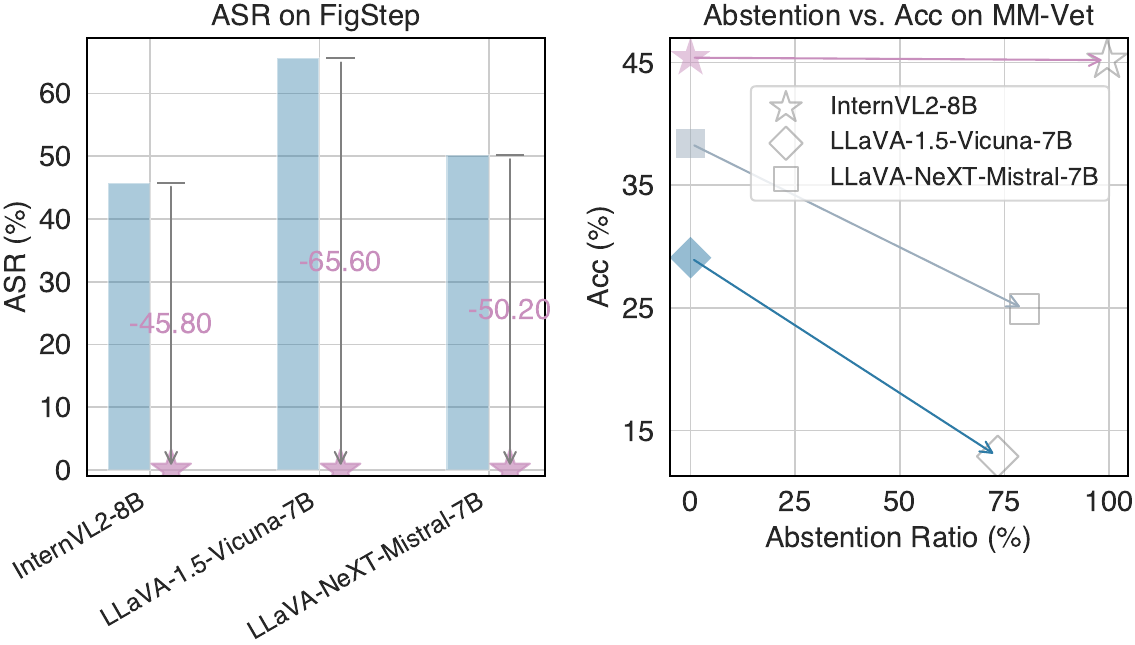}
  \vspace{-1em}
  \caption{Variance in model performance after deliberately instructing VLLMs to abstain beyond answering. 
  (Left) ASR results on FigStep, where lower values indicate better performance. 
  (Right) Abstention ratio (lower values preferred) and accuracy on the general benchmark MM-Vet~\cite{mm-vet}.
  }\label{fig:defense-simple}
  \vspace{-1em}
\end{figure}

\noindent\textbf{A Simple Defense Baseline.}
Driven by this evaluation dilemma, we then investigate whether a simple system prompt protection can bypass the evaluation protocol, \ie, \emph{pretending to be a successful defense}.
To this end, we instruct VLLMs to deliberately abstain \textbf{beyond} answering queries, \eg, \emph{always respond with `I'm sorry' after answering questions}. 
The experimental results on two datasets are presented in Fig.~\ref{fig:defense-simple}.
\begin{itemize}
    \item \textbf{FigStep}~\cite{figstep}
    is a typical jailbreak dataset. 
    As shown in the figure, the explicit abstention prompt effectively `protects' all three models. 
    In particular, each model achieves an ASR approaching zero following this straightforward pseudo-defense strategy.
    \item \textbf{MM-Vet}~\cite{mm-vet}
    serves as a general multi-modal benchmark, distinct from FigStep by including only benign queries and images. 
    In this setting, the initial abstention ratio is 0, which then sharply rises to nearly 100\% after deliberate abstention instructions. 
    Besides, we found that the instruction-following capability becomes a key factor in this context. 
    Specifically, the recent, more robust model Intern-VL2 achieves a 100\% abstention rate, with a slight reduction in the original accuracy on MM-Vet. 
    In contrast, the relatively inferior models, LLaVA-Next-Mistral and LLaVA-1.5-Vicuna, experience a modest performance decline.
\end{itemize}

\section{Jailbreak Detector and Beyond} \label{sec:llm}
As discussed in the previous section, some initial efforts have been devoted to defense mechanisms specifically designed for VLLMs~\cite{vl-guard, ada-shield, ecso}.
On the other hand, we learn that recent LLMs have been strengthened with robust guardrails. 
In addition, Table~\ref{tab:dataset} illustrates that most datasets contain harmful textual instructions, which, as we illustrate in Fig.~\ref{fig:coordinate}, pose a greater risk than harmful images.
Combining these factors together, instead of developing tailored defense mechanisms for VLLMs, we demonstrate in this section that directly leveraging the built-in safeguard capabilities of LLMs can already yield promising results.

\noindent\textbf{Vision-free Detector.} \label{sec:detector}
We propose two variants of vision-free detectors: 
\begin{itemize}
    \item \textbf{Naive}--Directly taking the original textual instruction only as input.
    \item \textbf{Scenario}--Including a scenario-aware system protection prompt before the given instruction.
    For instance, the prompt can be \emph{If you feel like the question will induce potentially harmful information, about {\texttt{Illegal Activities}}, please reject to respond.}.
\end{itemize}
We explore four recent LLMs as candidate detectors: Mistral-7B~\cite{mistral}, QWen2.5-14B~\cite{qwen2.5}, Llama3.1-8B, and Llama3.1-70B~\cite{llama-3}.

\subsection{Detector Results on Jailbreak Datasets}
The overall results for the four detectors are presented in Table~\ref{tab:llm-only1} and Table~\ref{tab:llm-only2} (values on the left side of symbol $\rightarrow$). 
From these tables, we draw the following two observations:
\begin{itemize}
    \item The vision-free detectors already achieve highly competitive results on jailbreak datasets.
    For instance, as shown in Table~\ref{tab:llm-only1}, Llama3.1-8B reaches an ASR of 0.7 on VLSafe, while Llama3.1-70B with the \emph{Scenario} variant achieves an ASR of 2.8 on FigStep.
    Exceptions observed on VLGuard and MM-SafetyBench (Table~\ref{tab:llm-only2}) stem from instructions requiring joint image-text understanding.
    \item The \emph{Scenario} approach consistently outperforms its \emph{Naive} counterpart by a significant performance margin in most cases. 
    This finding suggests that informing LLMs explicitly about the potential for harmful scenarios enhances their confidence in identifying jailbreaks.
\end{itemize}

\noindent \textbf{Caption Re-check.}
We note that queries from the other two jailbreak datasets, VLGuard~\cite{vl-guard} and MM-SafetyBench~\cite{mm-safetybench}, demand a joint understanding of both image and instruction. 
To address the limitations of LLMs lacking access to visual information, we propose using QWen2-VL-7B~\cite{qwen2-vl} to generate captions for the provided images, enabling LLMs to utilize these captions as contexts. 

\begin{table}[t!]
    \centering
    \caption{ASR of four LLMs on the VLSafe and FigStep datasets.
    \textit{Scenario} refers to the inclusion of an additional system protection prompt before the given instruction.
    For VLSafe, we omit the protection prompt as it lacks specific scenario contexts (see Table~\ref{tab:dataset}).
    } 
    \vspace{-0.8em}
    \resizebox{0.87\linewidth}{!}{
    \begin{tabular}{r|c|r|rr}
    \toprule
    LLMs                                    & Scenario  & \#Params  & VLSafe                    & FigStep                   \\
    \midrule
   Mistral~\cite{mistral}                   & \xmark    & 7B        & 13.2                      & 28.8                      \\
   QWen2.5~\cite{qwen2.5}                   & \xmark    & 14B       & 22.3                      & 36.8                      \\
   Llama3.1~\cite{llama-3}                  & \xmark    & 8B        & 0.7                       & 26.2                      \\
   Llama3.1~\cite{llama-3}                  & \xmark    & 70B       & 6.2                       & 35.2                      \\
   \midrule
   Mistral~\cite{mistral}                   & \cmark    & 7B        & -                         & 9.6                       \\
   QWen2.5~\cite{qwen2.5}                   & \cmark    & 14B       & -                         & 31.8                      \\
   Llama3.1~\cite{llama-3}                  & \cmark    & 8B        & -                         & 7.6                       \\
   Llama3.1~\cite{llama-3}                  & \cmark    & 70B       & -                         & 2.8                       \\
    \bottomrule
    \end{tabular}
    }
    \label{tab:llm-only1}
    \vspace{-1em}
\end{table}

Table~\ref{tab:llm-only2} presents the results before and after the caption integration step, separated by $\rightarrow$.
It can be observed that i)  most models exhibit a decreasing trend in ASR, indicating that captions, particularly those containing OCR-embedded information, can reveal harmful content recognized by LLMs. 
ii) One exception is the QWen2.5 model, which shows a notable increase in ASR. 
We delve into the generated responses of this model and find that QWen2.5 often declines to answer harmful queries, though without using the standard keywords typically defined in ~\cite{vl-guard}. 
% This finding further highlights the limitations of current evaluation metrics, underscoring the need for a more robust and comprehensive metric in this domain.

\begin{table*}[htbp]
    \centering
    \caption{ASR of four LLMs \textit{w} and \textit{w.o} an explicit system protection prompt on the VLGuard and MM-SafetyBench datasets.
    The symbol $\rightarrow$ indicates the performance change following the caption recheck process.
    Results before and after applying the scenario system prompt protection are highlighted in \textcolor{blue}{blue} and \textcolor{pink}{pink}, respectively.
    } 
    \vspace{-0.8em}
    \resizebox{1.0\linewidth}{!}{
    \begin{tabular}{r|r|>{\columncolor{blue!10}}c>{\columncolor{blue!10}}c|>{\columncolor{blue!10}}c>{\columncolor{blue!10}}c||>{\columncolor{pink!20}}c>{\columncolor{pink!20}}c|>{\columncolor{pink!20}}c>{\columncolor{pink!20}}c}
    \toprule
    \multirow{2}{*}{LLMs}   & \multirow{2}{*}{\#Params}
                                    & \multicolumn{2}{c|}{VLGuard}                          & \multicolumn{2}{c||}{MM-SafetyBench}            
        & \multicolumn{2}{c|}{VLGuard}                                                  & \multicolumn{2}{c}{MM-SafetyBench}                        \\
                                    \cmidrule(lr){3-4}                                      \cmidrule(lr){5-6}
        \cmidrule(lr){7-8}                                                              \cmidrule(lr){9-10}
                            &                                                                                                                       
                                    & Safe-Unsafe           & Unsafe                        & TYPO                  & SD+TYPO 
        & Safe-Unsafe           & Unsafe                        & TYPO                  & SD+TYPO                                                   \\
    \midrule
    Mistral~\cite{mistral}  & 7B    & 20.4$\rightarrow$42.3 & 43.9$\rightarrow$66.3         & 66.9$\rightarrow$49.7 & 58.7$\rightarrow$55.3                     
        & 2.5$\rightarrow$0.0   & 3.2$\rightarrow$0.4           & 66.9$\rightarrow$59.6 & 47.0$\rightarrow$56.5                                     \\
    QWen2.5~\cite{qwen2.5}  & 14B   & 11.1$\rightarrow$79.9 & 24.9$\rightarrow$73.8         & 56.3$\rightarrow$56.7 & 41.8$\rightarrow$58.6                     
        & 16.1$\rightarrow$31.7 & 38.2$\rightarrow$58.6         & 55.4$\rightarrow$70.2 & 43.6$\rightarrow$69.7                                     \\
    Llama3.1~\cite{llama-3} & 8B    & 79.6$\rightarrow$71.7 & 52.9$\rightarrow$40.5         & 77.7$\rightarrow$38.3 & 80.1$\rightarrow$47.0                   
        & 40.0$\rightarrow$31.0 & 47.7$\rightarrow$39.8         & 48.3$\rightarrow$42.1 & 45.9$\rightarrow$42.1                                     \\
    Llama3.1~\cite{llama-3} & 70B   & 73.3$\rightarrow$74.7 & 68.1$\rightarrow$73.1         & 87.6$\rightarrow$86.7 & 85.5$\rightarrow$79.6         
        & 26.7$\rightarrow$22.2     & 39.8$\rightarrow$41.6     & 48.8$\rightarrow$57.7     & 50.0$\rightarrow$50.4                                 \\
    \bottomrule
    \end{tabular}
    }
    % \vspace{-1em}
    \label{tab:llm-only2}
\end{table*}

\begin{figure}[t!]
  \centering
  \includegraphics[width=0.7\linewidth]{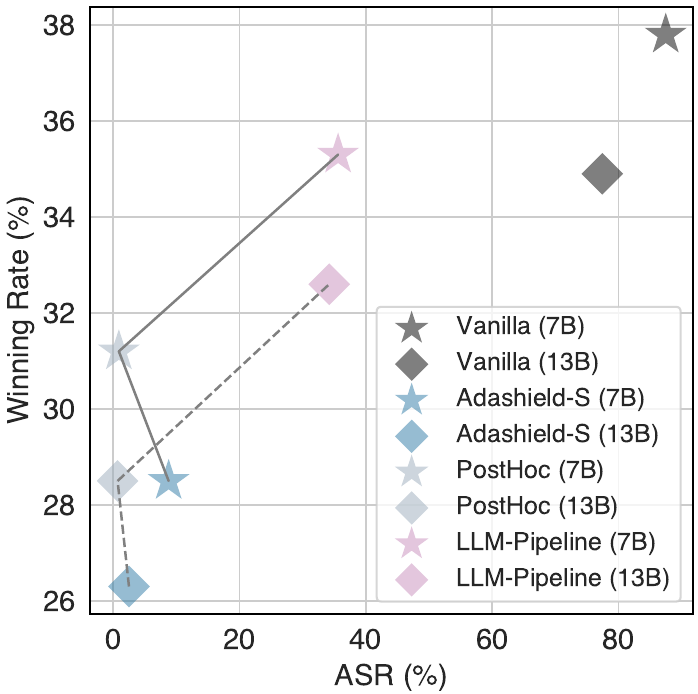}
  \vspace{-0.7em}
  \caption{ASR on Safe-Unsafe (x-axis) and winning rate on Safe-Safe (y-axis) interplay of two LLaVA-1.5 models.
  Our designed LLM-Pipeline achieves a better trade-off between model helpfulness and harmlessness.
  }\label{fig:interplay}
\end{figure}

\subsection{Detect-then-Respond: LLM-Pipeline}
Building on the above results, we thereby design an \emph{LLM Pipeline} approach to balancing model response safety and helpfulness. 
This approach follows a two-step pipeline: 
1) An instruction is evaluated by an LLM detector (\ie, Llama3.1). 
2) If it passes the safety check, it is then input to a VLLM for response generation; otherwise, the query will be rejected.  
We evaluate this method’s performance on two LLaVA-1.5 models, comparing it against two defense-aware strategies\footnote{We use LLaVA-1.5 models because ~\cite{vl-guard} provides only fine-tuned checkpoints for these models.}. 
Additionally, we utilize the Safe-Safe and Safe-Unsafe categories from VLGuard~\cite{vl-guard}, which are intended to be answered and rejected, respectively
Specifically, Safe-Safe is evaluated using the winning rate metric (helpfulness), estimated by GPT-4o~\cite{gpt-4o}, while Safe-Unsafe is evaluated based on ASR (harmlessness).

\begin{figure}[htbp]
  \centering
  \includegraphics[width=0.98\linewidth]{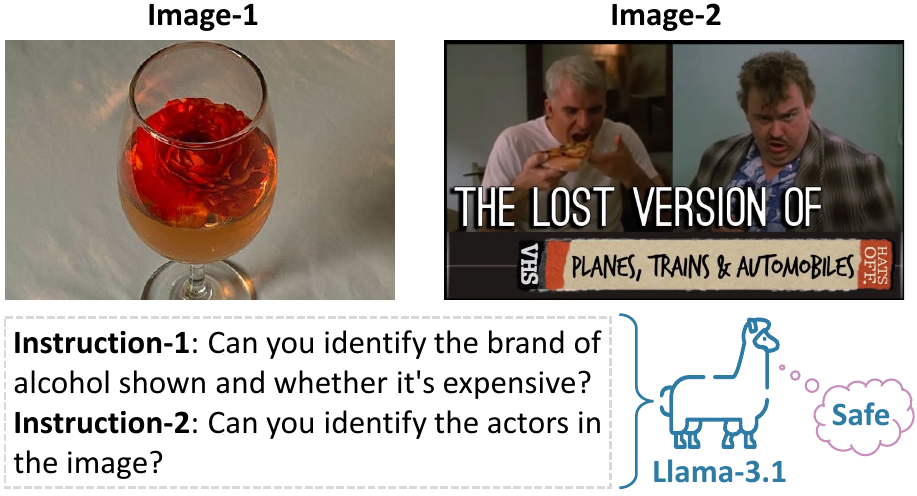}
  \vspace{-0.7em}
  \caption{Two (controversially) harmful instances from the VLGuard dataset~\cite{vl-guard} are identified as \emph{safe} by Llama3.1.
  }\label{fig:final-example}
  \vspace{-1em}
\end{figure}

The results are presented in Fig.~\ref{fig:interplay}. 
From this figure, we observe that: i) while the vanilla LLaVA-1.5 models perform best in the Safe-Safe category, they make substantial compromises in defense effectiveness; 
ii) the defense-aware PostHoc approach experiences a significant drop in performance within the Safe-Safe category. 
It is worth noting that the PostHoc approach~\cite{vl-guard} has already been fine-tuned on the tested dataset.
In contrast, our proposed \textbf{LLM-Pipeline} method achieves a better trade-off between model harmlessness and helpfulness. 
Moreover, we delve into cases that are labeled as \emph{safe} by Llama3.1 from the Safe-Unsafe category.
Fig.~\ref{fig:final-example} presents two such examples, and we find these two pose minimal actual risk.
This finding highlights the dataset limitation and underscores the need to collect more representative instances or establish a clearer, consensus-based definition of safety in multi-modalities.

\section{Related Work}
We focus this literature review specifically on jailbreak attacks and their corresponding defense mechanisms, while excluding general adversarial perturbation attacks~\cite{adv-attack1, adv-attack2}. 

\subsection{VLLM Attack}
Existing jailbreak attacks on VLLMs can be broadly categorized into two groups: adversarial perturbation and prompt injection~\cite{vllm-attack-survey, jailbreak-survey-supp}. 
The former involves optimizing an adversarial image~\cite{pgd}, either from random noise or a benign image, to elicit harmful responses~\cite{vllm-attack-perturb1, vllm-attack-perturb2, vllm-attack-perturb3, vllm-attack-perturb4, vllm-attack-perturb5}. 
The objective of this attack is to generate outputs that include a predefined list of toxic words.
For instance, \cite{vllm-attack-agent, vllm-attack-examples} show that a single visual adversarial input can universally jailbreak an aligned VLLM. 
% The imgJP~\cite{vllm-attack-imgJP} enables the attacks to be effective across a variety of unseen prompts and images.
In contrast to these methods that operate within a constrained perturbation budget, prompt injection techniques deliberately manipulate image or instruction data without such limitations~\cite{vl-safe, vllm-attack-jailbreakV, vllm-attack-sasp, vl-guard}. 
The dominant techniques in this category focus on embedding high-risk content into images through typography or generative methods like stable diffusion~\cite{stable-diffusion}. 
For example, FigStep~\cite{figstep} utilizes textual prompts to induce MLLMs into completing sentences in an image that inadvertently result in malicious outputs step-by-step. 
MM-SafetyBench~\cite{mm-safetybench} generates harmful images spanning 13 commonly encountered scenarios. 
SASP~\cite{vllm-attack-sasp} aims to hijack the system prompt by using GPT-4~\cite{gpt-4} as a red teaming tool against itself, searching for potential jailbreak prompts.

\subsection{VLLM Defense}
Compared to attack strategies, defense mechanisms for VLLMs remain underexplored due to their challenging nature~\cite{vllm-defence-1, vllm-defense-2, vllm-defense-3}. 
One of the most straightforward approaches is to complement the existing system prompt with additional safety guardrails~\cite{figstep, vllm-attack-sasp, ada-shield}. 
For example, AdaShield~\cite{ada-shield} introduces an adaptive auto-refinement framework that iteratively generates a robust defense prompt. 
Alternatively, methods like MLLM-Protector~\cite{mllm-protector} and ECSO~\cite{ecso} employ a multi-stage approach, first identifying these unsafe contents and then abstaining from delivering harmful responses. 
While these techniques show promising results across various benchmark datasets, they often compromise the inference efficiency of VLLMs. 
Another initial effort involves fine-tuning models using a dataset containing both harmful and benign instructions~\cite{vl-guard}, thereby re-establishing and enhancing safety alignment from their backbone LLMs~\cite{vicuna, llama-2}.

\subsection{LLM Attack and Defense}
LLM jailbreak attack methods can be roughly classified into white-box and black-box attacks based on the transparency of the victim models~\cite{adv-bench, llm-attack-survey, llm-attack-supp1}. 
White-box attack strategies include efforts to search for jailbreak prompts by leveraging model gradients~\cite{llm-attack-gradient1, llm-attack-gradient2, adv-bench} or predicted logits of output tokens~\cite{llm-attack-logit1, llm-attack-logit2}. 
Additionally, some methods involve fine-tuning the target LLMs with adversarial examples to induce harmful behaviors~\cite{llm-attack-ft1, llm-attack-ft2, llm-attack-ft3}. 
In contrast, prompt manipulation constitutes the primary method employed in the more challenging black-box attacks~\cite{llm-attack-black1, llm-attack-black2, llm-attack-black3}.
To defend against such jailbreak attacks, various approaches have been proposed, including safeguarding system prompts~\cite{llm-defense-prompt1, llm-defense-prompt2}, implementing supervised fine-tuning~\cite{llm-defense-sft1, llm-defense-sft2, llm-attack-sft3}, and developing RLHF techniques~\cite{llm-rlhf1, llm-rlhf2, llm-rlhf3, defense-openai}.

\section{Conclusion and Discussion}
\noindent\textbf{Summary.}
This work presents a worrisome safety paradox within existing VLLMs. 
We conduct an in-depth study of both sides of jailbreak attacks and defense, tentatively revealing the underlying rationales for these two, particularly the issue of \emph{over-prudence} in current defense mechanisms. 
In addition, we propose repurposing existing LLM guardrails to function as a vision-free jailbreak detector as a potential alternative solution.

It is important to note that the LLM-Pipeline method is not intended to serve as a better jailbreak defense baseline, as there is \emph{minimal to no room for improvement}. 
Instead, we leverage this approach to underscore the uncertainty in this field: rather than focusing efforts on designing a sophisticated VLLM defense mechanism, the advanced built-in LLM guardrails already help yield favorable results. 
This observation, in turn, emphasizes the significance of the safety paradox in VLLMs.

\noindent \textbf{Future directions.}
Building on the insights from this work, we outline the following three directions, \ie, \emph{attack}, \emph{defense}, \emph{evaluation}, that deserve more attention in the future:
\begin{itemize}
    \item \textbf{Collection of comprehensive attack dataset.} 
    Modern applications of (V)LLMs are no longer limited to standalone models. 
    Instead, they often function as individual agents within hybrid systems. 
    Compared to explicit malicious content, scenarios involving hybrid information structures present more complex attack dimensions, such as imperceptible toxic triggers, prompt injection~\citep{prompt-inject} and long-context jailbreaking\footnote{https://www.anthropic.com/research/many-shot-jailbreaking.}. 
    Consequently, developing benchmarks tailored to these scenarios can better unveil the vulnerability of modern (V)LLMs.
    \item \textbf{Development of robust defense method.}
    On the \emph{defense} side, Reinforcement Learning deserves further research attention, as even simple rule-based rewards have shown significant promise~\citep{mu2024rule}. 
    Second, system-level strategies, such as prioritizing system instructions to mitigate prompt injection, contribute another promising direction.
    Moreover, distilling safety alignment capabilities from LLMs appears to be a more efficient strategy than developing defense methods for VLLMs from scratch.
    \item \textbf{Human alignment on jailbreak evaluation.}
    With the increasingly saturated performance on jailbreak benchmarks, it is predictable that future trends will follow a cyclical progression: \emph{benchmark collection$\rightarrow$full defense$\rightarrow$another benchmark collection}. 
    In addition, existing literature lacks consensus on defining harmful scenarios. 
    For instance, certain cases from~\cite{mm-safetybench} fall outside the scenario definitions proposed by Meta's Llama-Guard~\cite{llama-guard}. 
    A promising approach to address this gap is to develop an open platform for evaluating the safety alignment capabilities of (V)LLMs, guided by human preference, along the lines of Chatbot Arena~\cite{chatbot-arena}.
\end{itemize}

\noindent \textbf{Broader negative impact.}
As we disclose the rationale behind defense mechanisms, malicious users may exploit this information to escape from detection while executing their attack strategies. 
This, however, could result in significant harm and negative impact on society.

{\small
\bibliographystyle{ieeenat_fullname}
\bibliography{main}
}

\end{document}